\documentclass[sigconf]{acmart}
\AtBeginDocument{%
  \providecommand\BibTeX{{%
    \normalfont B\kern-0.5em{\scshape i\kern-0.25em b}\kern-0.8em\TeX}}}

\usepackage{bm}
\usepackage{caption}
\usepackage{hyperref}
\hypersetup{breaklinks=true}
\usepackage{enumitem}
\usepackage{multirow}
\usepackage[super]{nth}
\usepackage{subfig}
\begin{document}

\title[]{Projecting Non-Fungible Token (NFT) Collections: A Contextual Generative Approach} 

\author{Wesley Joon-Wie Tann, \ Akhil Vuputuri, \ Ee-Chien Chang}
\affiliation{%
  \institution{Department of Computer Science, National University of Singapore}
  \country{}   }

\renewcommand{\shortauthors}{Tann et al.}

\begin{abstract}
Non-fungible tokens (NFTs) are digital assets stored on a blockchain representing real-world objects such as art or collectibles. An NFT collection comprises numerous tokens; each token can be transacted multiple times. It is a multibillion-dollar market where the number of collections has more than doubled in 2022.
In this paper, we want to obtain a generative model that, given the early transactions history (first quarter $Q_1$) of a newly minted collection, generates subsequent transactions (quarters $Q_2$, $Q_3$, $Q_4$), where the generative model is trained using the transaction history of a few mature collections. The goal is to use the generated transactions to project the potential market value of this newly minted collection over the next few quarters. A technical challenge exists in that different collections have diverse characteristics, and the generative model should generate based on the appropriate ``contexts'' of the collection.
Our method takes a two-step approach. First, it employs unsupervised learning on the early transactions to extract characteristics (which we call contexts) of NFT collections. Next, it generates future transactions of each token based on these contexts and the early transactions, projecting the target collection's potential market value. Comprehensive experiments demonstrate our contextual generative approach's NFT projection capabilities.

\end{abstract}



\maketitle

\section{Introduction} 
Non-fungible tokens (NFTs) broke into the mainstream and saw a boom in the past year~\cite{wang2021non}. 
The explosion in the popularity of NFT collections in the digital art space has garnered much interest among artists, private and institutional investors, and art galleries. 
While some collections are highly valuable, many others remain of little worth. 
As an immense number of artworks in the form of NFTs flood the market, it is difficult to determine the potential value of any particular NFT token or collection. Projecting its future value is very difficult~\cite{kugler2021non}. 
Nevertheless, major art galleries and investment institutions continue to pour billions into the sector~\cite{howcroft_2021,mozee_2021}, completely changing the landscape of fine art and traditional investments. 

\begin{figure}[tbp]
\centering
\resizebox{0.85\linewidth}{!}{%
  \subfloat[\#2087 (ETH 769)]{%
    \begin{minipage}[t]{0.32\linewidth}
      \includegraphics[width=\linewidth]{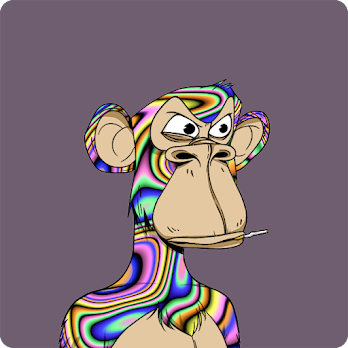}
    \end{minipage}}
  \hspace*{0.01\linewidth}%
  \subfloat[\#3749 (ETH 740)]{%
    \begin{minipage}[t]{0.32\linewidth}
      \includegraphics[width=\linewidth]{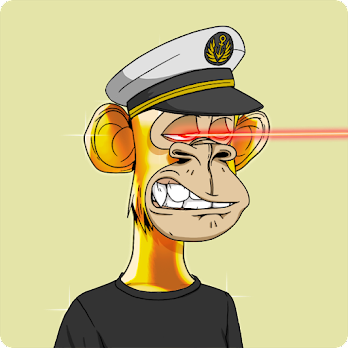}
    \end{minipage}}
  \hspace*{0.01\linewidth}%
  \subfloat[\#4580 (ETH 666)]{%
    \begin{minipage}[t]{0.32\linewidth}
      \includegraphics[width=\linewidth]{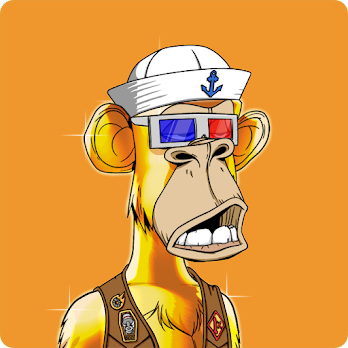}
    \end{minipage}}
}


\resizebox{0.85\linewidth}{!}{%
  \subfloat[\#7608 (ETH 0.069)]{%
    \begin{minipage}[t]{0.32\linewidth}
      \includegraphics[width=\linewidth]{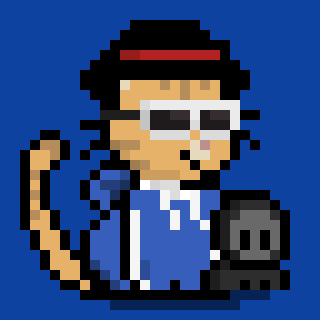}
    \end{minipage}}
  \hspace*{0.01\linewidth}%
  \subfloat[\#1470 (ETH 0.4)]{%
    \begin{minipage}[t]{0.32\linewidth}
      \includegraphics[width=\linewidth]{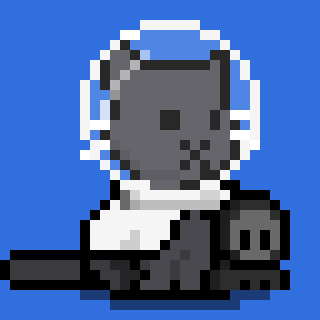}
    \end{minipage}}
  \hspace*{0.01\linewidth}%
  \subfloat[\#5037 (ETH 0.19)]{%
    \begin{minipage}[t]{0.32\linewidth}
      \includegraphics[width=\linewidth]{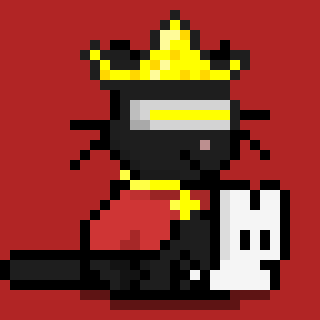}
    \end{minipage}}
}
\caption{The top NFT tokens of two collections with contrasting values, measured by market capitalization. While the \textit{Bored Ape Yacht Club (BAYC)} collection (upper row) consists of top transactions of around ETH 700 $\simeq$ USD 1M, 
the top transactions of the \textit{NEKO} collection (lower row) are usually less than ETH 1 $\simeq$ USD 1550. The reported last sale price of tokens (accurate as of Jan 20, 2023). 
} 
\label{fig:compareNFTs}
\end{figure}

An increasingly large number of new NFT collections are entering the market~\cite{zhuang_2022}. While each collection has a distinct artistic theme, the tokens of any particular collection can look very similar (see Figure~\ref{fig:compareNFTs}).
We wonder if there is an effective method that allows us to leverage the different characteristics of various established collections to generate transactions of new NFTs.

Given a newly minted NFT collection that has been in the market for a few months, our contextual generative approach aims to generate its future transactions, allowing us to project the potential market value. However, we observe that because the NFT market behaves similarly to a limited collectible market, transactions of each token are few and far between; this peculiar characteristic of NFT transactions is challenging for machine learning. Mainly, models trained with different collections result in disparate outcomes, while a model trained on multiple collections merely ``averages'' the results. 
%
Hence, we take a two-step contextual generative approach where 
\begin{enumerate}
    \item the semantic context of each NFT collection is first distilled and extracted, then 
    \item this auxiliary contextual information guides our generative method 
\end{enumerate} 
to generate future transaction data, resulting in contextualized projections.

Since there are thousands of tokens in any particular NFT collection, these aggregate transactions capture rich market supply and demand information, which can help us better understand the economics of NFTs. Never before have such alternative investment data been so publicly and readily available. Consequently, we ask this meaningful question:
\vspace{1mm}
\begin{center}
\noindent\parbox{0.85\linewidth}{%
\emph{Can we effectively distill contextual information from various NFT collections and leverage them to generate representative transaction data of unobserved NFTs, reflecting the potential value of new collections?}
}%
\end{center} 
\vspace{1mm}
If we can reasonably answer this question, embracing the recent advances in deep conditional machine learning, we would have the means of addressing this challenging market projection puzzle using financial data that was once closely guarded but has now become widely available. Moreover, it is a markedly novel investment class that provides insights into the world of alternative investments, which was once the exclusive domain of the privileged.

Existing models for forecasting the value of digital collectibles recorded in the decentralized public ledger, particularly NFTs, are mainly limited to two categories. 
Currently, these models are predominately based on either (1) traditional financial asset pricing methods~\cite{dowling2022non,kong2021alternative,schnoering2022constructing,goldberg2021economics} or (2) approaches studying external factors (e.g., the underlying cryptocurrencies, Twitter influence, visual features) that affect NFT prices~\cite{nadini2021mapping,dowling2022fertile,kapoor2022tweetboost,vanpredictive}.
While the traditional financial pricing approach performs empirical analysis such as statistical analysis and economic forecasts on standard financial indicators, analyses of external factors focus on how much of a correlation exists between NFT prices and the studied factors. However, these approaches do not fundamentally consider the underlying structure of NFT transactions, supply and demand relationships, and how it impacts NFT prices.

In contrast, we directly address the question of whether market behaviors captured in transaction series indicate NFT prices. Our proposed approach leverages conditional generative models~\cite{mirza2014conditional,NIPS2019_8415,guo2020recurrent} and LSTM networks~\cite{hochreiter1997long} for future transaction series generation.
The first two stages of development of NFT collections can be broadly classified as the early stage (initial 3 months) and the growth stage (next 9 months). 
First, we analyze each NFT in its early stage. 
Then, constructing a context vector $C_i$ through unsupervised learning for each collection $i$ based on its first quarter ($Q_1$) transaction series $T_i$ provides some particular context of the collection. The transaction series $T_i$ consists of the daily values and transaction count of each token in the collection.

Next, employing these contexts as conditions, it is concatenated with the transaction series and fed as inputs $(C_i, T_i)$ to our model. 
The model then learns from the contexts and transactions of established NFT collections. Given a new NFT collection in its early stage, it effectively generates future transaction series of the new collection. Lastly, we perform a \textit{step-transform} procedure on the generated series, following the piecewise constant series that characterizes each token transaction series. Such a generative modeling approach produces unobserved future series (See Section~\ref{subsec:approach} for details).

Our experiments are performed on real-world collections in the NFT market. We used transaction data from five collections for training and evaluated the proposed generative approach on five other collections (see Tables \ref{tab:predictperf} and \ref{tab:azuki_analysis}). 
In the experiments, we first train our model on the five training collections of various total market capitalizations. 
Then, from each collection, we construct transaction series of size equal to the number of tokens and length of 365 to reflect the early and growth stages. Every collection has its associated context vector that is characteristic of its transactions in the early stage (see Figure~\ref{fig:embedding}).
For example, for the training collection BAYC, there are 10,000 tokens. The transaction series, size $[10,000 \times 365 \times 2]$, will have its 6-dimensional context vector concatenated at the start to make up the inputs to the model. Therefore, the model learns the transaction characteristics and series of various collections at different developmental stages.

Next, given a new NFT collection in its first quarter, the model derives its context through the unsupervised learning method and uses it to generate the series for the next few quarters of this new collection. While we use the \textit{PCA} method~\cite{jolliffe2002principal} here, any other unsupervised learning technique is applicable. 
Empirical results show that our approach can accurately generate future series, thereby projecting the growth of NFTs (see Table~\ref{tab:predictperf}). Furthermore, we set up baseline models trained on individual training collections and an aggregate unconditional model that does not consider the context information. The results show that our approach significantly outperforms the baseline comparisons, demonstrating the power of contextual generative modeling NFT transaction series.

\vspace{1mm}
\noindent{\bf Contribution.} 
\begin{enumerate}
\setlength\itemsep{0.5em}

\item We identify the significance of numerous NFT collections in the market, each with different characteristics, and their corresponding transaction series to estimate the potential value of such alternative digital assets. 


\item We introduce a two-step contextual generative approach that directly leverages the diverse characteristics of NFT collections to generate future transactions.

    \begin{enumerate}
        \item First, using unsupervised learning, we distill key contextual information from early transactions ($Q_1$) of various collections.
        
        \item Second, given these contexts and early transactions, the approach generates future transactions ($Q_2$, $Q_3$, $Q_4$) of tokens over time, projecting the target collection’s potential market value. 
    \end{enumerate}

\item We present experimental results on real-world NFT collections to support our proposed approach. Our analysis demonstrates that the approach is able to generate future transactions that project NFT growth, outperforming baseline methods.

\end{enumerate}

\section{Background and Related Work} 
\label{sec:background}
In this section, we explain the concept of non-fungible tokens and discuss NFTs in the domain of blockchains and cryptocurrencies. Next, we describe the NFT transaction series and valuation methods. Finally, we investigate existing contextual deep generative models.


\begin{figure*}[!tbp]
\centering
\resizebox{0.85\linewidth}{!}{%
    \begin{minipage}[c]{.5\linewidth}
        \centering
        \includegraphics[width=\linewidth]{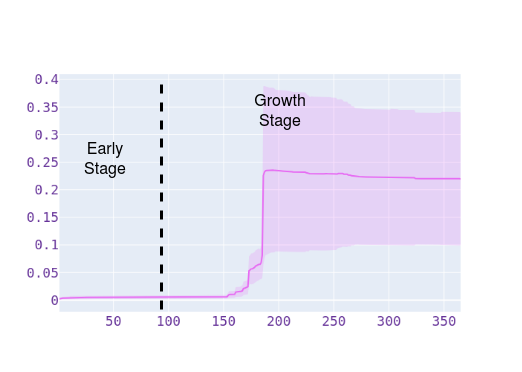}
        \caption*{Neko: first-year growth}
    \end{minipage} \hfill \hspace{5em} 
        \begin{minipage}[c]{.5\linewidth}
        \centering
        \includegraphics[width=\linewidth]{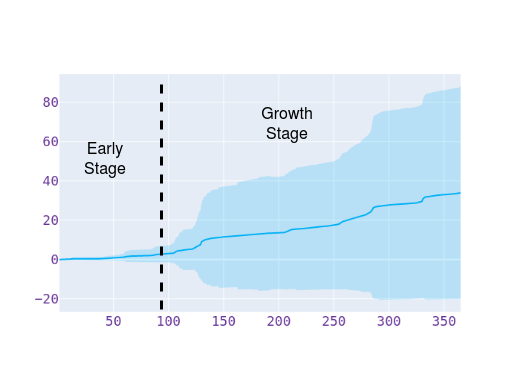}
        \caption*{BAYC: first-year growth}
    \end{minipage}
}
\caption{Mean and variance of Neko (left) and BAYC (right) collections in their first year, where the daily transaction values (Y-axis) are plotted against the number of days (X-axis). As shown, the early stages are indicative of potential value. The mean values of Neko plateau around ETH 0.22 in the growth stage with a converging variance, while BAYC continues to grow with increasing variance. 
}
\label{fig:visualnetwork}
\end{figure*}
    
\subsection{Non-Fungible Tokens}
Non-fungible tokens (NFTs) are digital assets that exist in smart contracts of the Ethereum~\cite{wood2014ethereum} blockchain. It can represent real-world objects such as art, music, in-game items and videos, collectibles, and even real estate. 
NFT differs from standard cryptocurrencies~\cite{phillip2018new} in its fundamental property. While a cryptocurrency (e.g., Bitcoin~\cite{nakamoto2008bitcoin}) is identical to another, therefore serving as a medium of exchange, NFTs are uniquely identifiable.
Specifically, an NFT is a unit of data stored on a blockchain that certifies digital representations of physical assets to be unique. Initially introduced in the ERC-721 standard~\cite{william2018erc} for representing ownership of non-fungible tokens, that is, where each token is unique. Besides providing a form of digital ownership certification, an NFT introduces the proof of assets right from inception~\cite{nadini2021mapping}.

Artists and content creators alike can now easily prove their ownership of digital assets.
Although, in essence, NFTs represent little more than code, the codes to a buyer have ascribed value when considering its comparative scarcity as a digital object. It secures the selling prices of these intellectual properties that may have seemed unthinkable for non-fungible virtual assets~\cite{wang2021non}. Originally NFTs were part of the Ethereum blockchain, but increasingly more blockchains have implemented their versions of NFTs~\cite{locke_2020}.

\subsection{Transaction Series} 
NFT marketplace activities can be categorized into four types: Listings, Transfers, Bids, and Sales~\footnote{Further descriptions of various activity types can be found at \url{https://support.opensea.io/hc/en-us}}. A \textit{listing} is created when an owner of a wallet containing the NFT makes it available for sale in the market. 
An NFT \textit{transfer} is an activity that enables an easy way of transferring tokens to another wallet, which could belong to friends, fellow community members, or perhaps just another wallet of the same owner. As NFT marketplaces resemble fine art auction houses, they operate live bidding auctions where people \textit{bid} for any particular NFT of their interest, and the highest bid wins. Finally, a \textit{sale} is an actual transaction between the buyer and seller. 
%
By taking these transactions of each NFT token, we construct transaction series that capture meaningful price behaviors among tokens. 
For example (see Figure~\ref{fig:visualnetwork}), the early stage transactions of the BAYC collection distinctly vary from the NEKO collection, which is indicative of their characteristics at a later growth stage.

\subsection{NFT Valuation and Pricing}
Existing NFT pricing and asset valuation can be broadly categorized into two branches. In one branch, the approaches~\cite{dowling2022non,kong2021alternative,schnoering2022constructing,goldberg2021economics} originate from the finance discipline. NFTs are viewed as alternative investments and studied with either supply-and-demand models or empirical methods such as regression models and time-series economic forecasts.
The other branch~\cite{nadini2021mapping,dowling2022fertile,kapoor2022tweetboost,vanpredictive} takes a different approach, studying whether NFT prices are affected by external factors (e.g., the underlying cryptocurrencies, Twitter influence, visual features) and how much of a correlation between the factors.

Nadini et al.~\cite{nadini2021mapping} showed in linear regression analyses that visual features of NFTs have significant predictive power on future primary sale prices, resulting in regression coefficients (Adjusted R-squared $R^2_{adj} \in [0.40, 0.50]$). Another work~\cite{dowling2022fertile} explores the relatedness of NFT pricing that is driven by cryptocurrencies. 
Their spillover index and wavelet coherence analysis indicate some co-movement between the NFT and cryptocurrency markets with limited volatility transmission effects, suggesting that cryptocurrency pricing behaviors might help understand NFT pricing patterns. 

As for other works~\cite{kapoor2022tweetboost,vanpredictive}, they studied the effects of social media on NFT pricing. In particular, Twitter and its social media features were proposed as a predictive factor for NFT prices. The presented results show that social media features (e.g., count of user membership lists, number of likes, retweets) have important predictive value. 
However, there has not been a study on the effects of NFT transaction series and their predictive power over the future valuation and success of new NFTs.

\subsection{Contextual Generative Methods}
Semantic contexts have been leveraged for both conditioning deep graph generative models~\cite{NIPS2019_8415,li2018learning,NEURIPS2019_b0bef4c9} and natural language generation~\cite{guo2020recurrent,fedus2018maskgan}. 
These conditional models either add the semantic context information in the inputs or latent representation during the learning process. However, to the best of our knowledge, the contexts are hand-crafted according to some semantic meaning in each specific domain.
On the contrary, we employ unsupervised learning to derive representative and meaningful contexts. Moreover, there are no existing works on the generation of NFT transactions.

\section{Projecting NFT Collections} 
\label{sec:nftsuccess}
In this section, we first formulate the novel problem of generating future transaction series to project the success of new NFT collections in the market. We then provide a motivating example to highlight the issue and offer a practical instance of such projections. 
Next, we present the approach, employing conditional generative modeling~\cite{mirza2014conditional,NIPS2019_8415,guo2020recurrent} and LSTM~\cite{hochreiter1997long} networks for generating future transaction series.

\subsection{Problem Formulation}
\label{subsec:formulate}
The development of an NFT in the marketplace can be broadly divided into three key stages: (1) early stage, (2) growth stage, and (3) late stage. These three phases can be loosely defined as: 
\begin{enumerate}
    \item Early stage (3-month period after mint): This stage begins when the NFT is made available for sale and introduced to the markets for a target community. The market is exposed to the new piece and recognizes its potential value during this stage. 
    \item Growth stage (an extended period after the early stage): This stage begins when a majority of the community is aware of the new NFT. The market starts to trade the NFT up to its perceived value—usually over a 9-to-12-month period. This is the period that attracts investments and sees very significant price appreciation. 
    \item Late stage (after growth stage): This stage is when most of the NFT potential growth is priced in, and the price movements have generally stabilized, barring unexpected idiosyncratic risks to the NFT or systemic shocks to the markets. 
\end{enumerate}
In our problem, we only consider the first two stages of development because our goal is to project the future success of new NFTs. Based on the past nature of NFT markets and demand, there is much more potential for growth when an NFT is at its early stages.

\begin{figure}[!tbp]
    \centering
    \includegraphics[width=.8\linewidth]{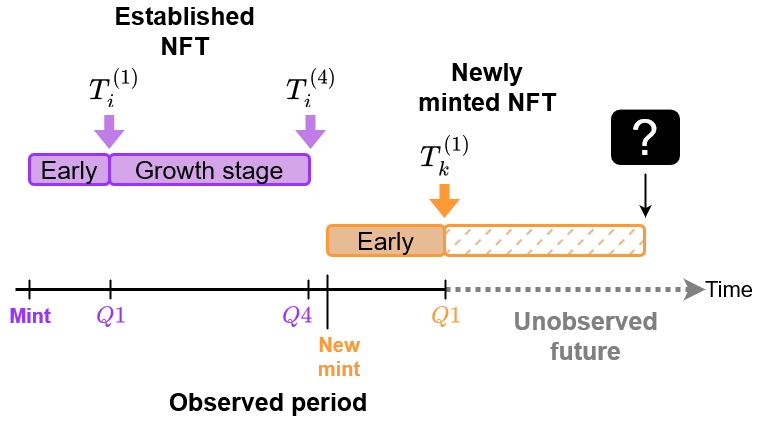}
    \caption{Projecting the future transaction series $T_k$ of a newly minted NFT by leveraging the observed transaction series of established NFTs $T_i$ in the growth stage. 
    }
    \label{fig:nfttimeline}
\end{figure}

\vspace{1mm} 
\noindent\textbf{Proposed Problem.} 
We focus on the novel problem of projecting the success of newly minted NFTs (see Figure~\ref{fig:nfttimeline}). Here, \textit{success} is defined as the popularity of distinguished NFTs, reflected by the value of transactions in the market. 
As forecasting market prices and conquering the world of investing is an age-old riddle, many have relied on conventional approaches that hinged on tried-and-tested methods such as statistical time series analysis and even more advanced sequence machine learning techniques to forecast price movements.
However, NFTs, unlike the stock markets where numerous daily transactions help with value discovery, resemble the fine art market with transactions few and far between. We depart from the convention and ask the following question: Can we model the characteristics of a new NFT's transaction series based on established NFTs in the market to project its future growth potential?
\vspace{1mm}
In other words:
\begin{center}
\noindent\parbox{0.91\linewidth}{%
    {\em Is it possible to learn the established NFTs' contextual embeddings $C_i$ and transaction series $T_i$ to accurately generate a new NFT's future series $T_k$, which is a projection of its potential success?
    }
}%
\end{center} 
\vspace{1mm}

\noindent If we can embrace the generated future transaction series for projection and put forward a method to answer this question adequately, we would have the means of addressing this enduring market forecasting puzzle from another angle that could provide insights and allow us a peek into the future market.

Hence, we formalize this novel problem as the following. Given an observed set of established NFTs passed their growth stages, transaction series $\mathcal{T} = \{ T_1, T_2, \dots, T_m \}$ of $m$ different NFT collections, where each $T_i = (\mathbf{x}^{(t)}_i, \dots, \mathbf{x}^{(t+n)}_i)$ 
corresponds to a particular transaction series of an NFT collection. Each sample from the series is described by two vectors, (1) the $n$ daily values of tokens and (2) the transaction number of the daily values.
For example, in the first quarter ($Q_1$) of an NFT collection BAYC, its transaction series would contain 91 $\mathbf{x}$'s for all 10,000 tokens in the collection. Each $\mathbf{x}$, size $[91 \times 2]$, contains a vector of token values (based on last transaction) and a vector of the transaction number. Hence, the BAYC Q2 series is a $[10000 \times 91 \times 2]$ array.

As each new NFT collection gets minted, another transaction series $T_k = (\mathbf{x}^{(t)}_k, \dots, \mathbf{x}^{(t+n)}_k )$ will be observed as this NFT is transacted among users in the early stage.
Since each transaction series can be examined as early as its first quarter $Q_1$ (early stage) of development, which is indicative of its potential success, we construct a context vector $C_i$ for each series based on its $Q_1$ transaction series $T_i$, which describes some particular transaction behavior contexts of the collection (see Section~\ref{sec:eval} for NFT statistics).

In this work, we aim to model transaction series characteristics of established NFTs in different stages of their development and project the potential of new NFTs by generating their future series and measuring their market growth potential. By training model $\mathcal{M}$ on a set of transaction series with their corresponding conditions ($\mathcal{D} = \{ C_i, T_i \}^m_{i=1}$), we aim to achieve two objectives. (1) Given a condition $C_i \in \mathcal{C}$ of observed transaction series, generate series $T_i$ that resemble the characteristics of those in the training set $\mathcal{D}$. (2) Given a condition $C_k \notin \mathcal{C}$ of unobserved series, generate future series $T_k = (\mathbf{x}^{(t)}_k, \dots, \mathbf{x}^{(t+n)}_k)$ of new NFTs that projects of their growth, providing insights into the unknown market conditions.

\vspace{1mm} 
\noindent\textbf{Notations.}
Here is a summary of the notations used.
\begin{itemize}[leftmargin=1.2cm,labelsep=0.0cm]
\setlength\itemsep{.2em}

\item[$\mathcal{T}$: \ \ ] Set of all observed transaction series of \textit{established NFTs}
\item[$T_i$: \ \ ] A particular $i$-th collection series from the set of observed established NFTs set
\item[$T_k$: \ \ ] Transaction series of a \textit{newly minted NFT}

\item[$\mathbf{x}^{(t)}_i$: \ \ ] The values of all tokens in any established collection $i$ at a particular time step $t$
\item[$\mathbf{x}^{(t+n)}_i$: \ \ ] Value of tokens from a collection $n$ days later
\item[$\mathbf{x}^{(t)}_k$: \ \ ] The values of all tokens in a new collection $k$ at a time step $t$
\item[$t$: \ \ ] Each time step
\item[$n$: \ \ ] Number of time steps
\item[$C_i$: \ \ ] Context vector of transaction series, describing some contexts of the corresponding collection
\item[$\mathcal{D}$: \ \ ] Training set that includes all the observed transactions and their respective condition vectors
\item[$\mathcal{M}$: \ \ ] Generative model for the transaction series generation 
\end{itemize}

\subsection{Motivating Scenario}
\label{subsec:example}
Let us consider an actively managed investment fund (e.g., hedge fund, sovereign fund) that aims to enter the NFT markets because it has identified NFT's profit potential or wants to diversify its portfolio with this alternative asset class. The fund chooses to invest in newly minted NFTs instead of established ones as eminent NFTs have already seen a vast price run-up, and it expects there to be much more growth potential in new NFTs. 

Since NFTs in the digital space are publicly available, it is reasonable to assume that the investment fund can access various established collections' data (e.g., wallet addresses, token IDs, and transactions). The fund could then hire experts to survey the difference between the collections and how their transactions vary accordingly and base its investment decisions on such ``qualified'' opinions up to a confidence level. On top of such a qualitative report, it has to provide another form of market analysis and deeper quantitative reasoning that measures the potential growth to convince key stakeholders of the fund.

Given such a scenario, can we distill the contextual information of established NFTs and leverage the contexts to generate future transactions of a newly minted NFT, projecting an accurate quantitative measure of the target NFT collection's potential market value? 
In fact, market transactions of various collections are rich with contextual information. Hence, we investigate whether such contexts $C_i$ and transaction series $\mathcal{T}$ of established NFTs can be used for generating future transactions $T_k$ of new NFTs, projecting their potential market value.

\subsection{Generating Future Transactions} 
\label{subsec:approach}
We present the approach employed to learn from observed transaction series of NFTs and generate unknown series of new NFTs resembling the future, projecting their success and growth. It harnesses conditional modeling and deep LSTM networks for conditional series generation. 
%

\begin{figure}[H] 
    \centering
    \includegraphics[width=0.8\linewidth]{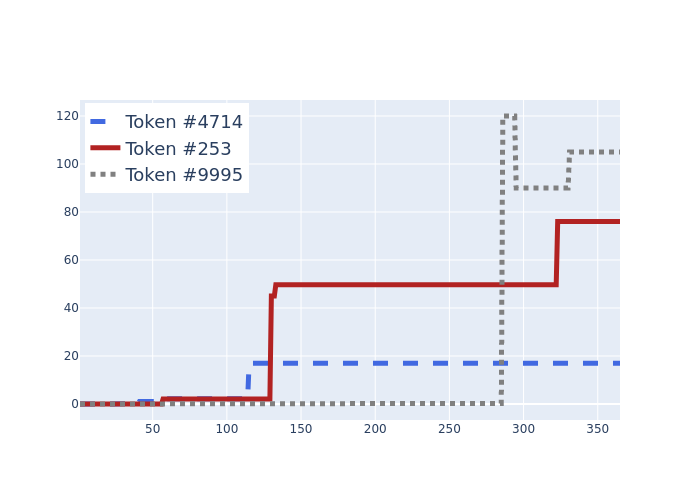}
    \caption{Sample token series from BAYC collection, showing the piecewise constant characteristics of NFT transactions.
    }
    \label{fig:steptrain}
\end{figure}

NFT transaction series is an accumulation of sequences; each sample is a sequence of a token, where an element is the last transacted price and its transaction count.
Every sample of the transaction series, reflecting the daily value of each token based on the last transaction price, resembles a step function. For example, tokens \#4714, \#253, and \#9995 of the BAYC collection are all transacted four times, respectively. These tokens have the following series (see Figure~\ref{fig:steptrain}).
NFT transactions are both irregular and infrequent. Most of the tokens have zero number of transactions in the first year. However, most tokens in popular collections, such as BAYC and Hashmasks have 2--6 transactions in the first year (see Appendix for histogram analysis).

Given that the samples are piecewise constant functions, this poses a fundamental challenge to the generation of transaction series as existing sequence models are designed for smooth and continuous functions. To meet this challenge, we design a method that is able to represent the distribution of such piecewise constant functions and generate future steps from the estimated distribution---in particular, modeling the successive relationship between transaction values and counts from observed NFT tokens.

\vspace{1mm} 
\noindent\textbf{Conditional LSTM Generative model.}
LSTM networks~\cite{hochreiter1997long} are widely adopted in sequence modeling, such as natural language processing~\cite{graves2013generating,vaswani2017attention}, where there are complex sequences with a long-range structure. 
We propose a direct conditional method that leverages the sequence modeling power of the LSTM, while retaining its architecture. By conditioning the model with additional information, we control the generative process. Our model is conditioned on the transaction profiles of collections in the first quarter $Q_1$. First, we perform a principal component analysis (PCA)~\cite{jolliffe2002principal} on the $Q_1$ transaction series of all training collections.

Next, we take the clustering centers~\cite{lloyd1982least} (means of the PCA transformed data sample points) from each collection. These $Q_1$ transactions are encoded to a sequence of 6-dimensional context vectors $C_i \in \mathbb{R}^6$. Since components in the PCA could either be positive or negative, depending on the direction, we normalize them between $a$ and $b = 3$ for training the LSTM model. 
In order to normalize the context values, we first concatenate the 6D vectors into an array $\mathcal{C}$. Next, we add the absolute minimum value to the array $\bf{C} = \mathcal{C} +  \mid{C}_{min}\mid$ and apply the following normalization formula:
\begin{equation}
    (\bf{C} - \bf{C}_{min}) / (\bf{C}_{max} - \bf{C}_{min}) * (b-a) + a
\end{equation}
where $\bf{C}_{min}$ and $\bf{C}_{max}$ are the minimum and maximum values in $\bf{C}$, respectively; the lowest value $a = 1$ and highest value $b = 3$.
It results in context vectors with values between 1 and 3. As these contexts are inputs to our LSTM models, it is important that the values are positive to improve the training.

With these context vectors as conditions, we concatenate them with the $Q_1$ transaction series as input to the LSTM model. 
At any particular timestep $t$, an LSTM cell has three gates: input $i_t$, forget $f_t$, and output $o_t$. The architecture below:
\vspace{1mm}
\begin{align}
\setlength{\itemsep}{7pt}%
  &\begin{aligned}
    i_t &= \sigma(W_{i}[(C_i, x_t), \ h_{t-1}] + b_i)	
  \end{aligned}\\
  &\begin{aligned}
    f_t &= \sigma(W_{f}[(C_i, x_t), \ h_{t-1}] + b_f) 
  \end{aligned}\\
  &\begin{aligned}
    o_t &= \sigma(W_{o}[(C_i, x_t), \ h_{t-1}] + b_o) 
  \end{aligned}\\
  &\begin{aligned}
    h_t = o_t \odot \tanh(c_t) 
  \end{aligned}\\
  &\begin{aligned}
    c_t &= f_t \odot c_{t-1} + i_t \odot \tilde{c}_t
  \end{aligned}\\
  &\begin{aligned}
    \tilde{c}_t = \tanh(W_c[(C_i, x_t),  h_{t-1}] + b_c) 
  \end{aligned}
\end{align}
where $C_i$ is the condition of the $i^{th}$ NFT collection, $x_t$ is the input, $h_{t-1}$ the LSTM cell output from the previous timestep, $W$ and $b$ are the respective weights and biases, $\sigma(\cdot)$ is the sigmoid function, $\tanh(\cdot)$ is the hyperbolic tangent function, and $\odot$ denotes element-wise product.
The output of the model is projected to a 2-dimensional vector, predicting the next timestep in the series. A sliding window moves accordingly and is concatenated with the context condition to generate the future timesteps. This procedure is repeated until the model generates the specified number to timesteps into the future (see Appendix for implementation details). 

Lastly, as the generations of LSTM return in a smooth curve and the nature of NFT transaction series are piece-wise constant, we perform an additional step-transform procedure. Based on the generated transaction counts, we take the first value of each count and keep all the following steps of the same count to have the same value. It results in our projection following a step function and resembles real NFT transactions.

\section{Evaluation} 
\label{sec:eval}
Our evaluation is performed on five real-world test datasets of NFT collections in the open market, \textit{Chubbies}, \textit{CryptoTrunks}, \textit{VeeFriends}, \textit{MAYC}, and \textit{Azuki}. We conduct thorough experiments to demonstrate that the contextual generation of future transaction series can project the growth of NFTs. All code used in our experiments is available on GitHub~\footnote{\url{https://github.com/}}.

\subsection{NFT Collections} \label{subsec:data}
We train our contextual generative approach with five market-traded NFT collections~\footnote{\url{https://opensea.io/rankings?sortBy=total_volume&category=art}}.
These are tokens that operate on the Ethereum~\cite{wood2014ethereum} blockchain, which is the foremost blockchain for trading NFTs. It is a public blockchain network. Hence, all the Ethereum transaction data is freely available, and we access the data through a blockchain explorer---Etherscan~\footnote{\url{https://etherscan.io/apis}}.

\vspace{1mm} 
\noindent\textbf{Training Collections.}
The five NFT collections are ranked according to their success and popularity, measured by total volumes (ETH) as of Jan 2023, listed below:
\vspace{1mm}
\begin{itemize}
\addtolength{\itemindent}{0.9cm}
\setlength{\itemsep}{3pt}%
    \item[($X_1$)] \emph{Bored Ape Yacht Club (BAYC)} --- ETH 710K 
    \item[($X_2$)] \emph{Hashmasks} --- ETH 43K 
    \item[($X_3$)] \emph{Polychain Monsters} --- ETH 3.5K 
    \item[($X_4$)] \emph{NEKO Official} --- ETH 672 
    \item[($X_5$)] \emph{EtherThings} --- ETH 599 
\end{itemize}
\vspace{1mm}
These are limited NFT collections (see Appendix for description). In each collection, some NFT tokens have turned out to be highly valued, while others have not seen any success, even though they are from the same collection.

First, we retrieve the first-year data of each of the collections, starting from their inception dates. Then, we extract the transactions of all tokens from a collection. Using these transaction values and their transaction count, we form the quarterly ($Q_1$--$Q_4$) transaction series, where $Q_1$ is the early stage and $Q_2$--$Q_4$ is the growth stage. 

For example, there are 10K tokens in the BAYC collection. Each token has a transaction series of 365 days. 
For a token with two transactions in the first quarter (91 days) with the first transaction value of $ETH \ 2$ and the next transaction of $ETH \ 159$ (e.g., $[(0,0), (0,0), (2,1), (2,1), \dots, (159, 2), (159, 2)]$), it has a transaction series reflecting the values and transaction count during the period.

Based on these series, we compute important transaction statistics of each quarter, thereby retrieving their representative market characteristics. A summary (see Table~\ref{tab:quintilestats}) of the market statistics for four quarters shows that the characteristics of transaction series vary from one quarter to the next.
    \begin{table}[!htbp]
\centering
\caption{Market statistics of the first four quarters of each training collection at the different stages---early and growth.}
\resizebox{0.99\linewidth}{!}{%
    \begin{tabular}{|c|c|ccccc|}
    \hline
    \textbf{Collection} & \textbf{Quarter} &\textbf{Market Cap} &\textbf{High} & \textbf{Low} & \textbf{Mean} & \textbf{Change (\%)} \\ \hline
    \multirow{4}{*}{\begin{tabular}[c]{@{}c@{}}\textit{BAYC} 
    \end{tabular}}      
    &$Q_1$ &24,252.51   &105.00     &0.0001  &0.98   &-- \\ 
    &$Q_2$ &117,718.68  &769.00     &0.0001  &8.75   &385.39 \\
    &$Q_3$ &196,391.34  &769.00     &0.0000  &16.45  &66.83 \\
    &$Q_4$ &307,509.66  &1,080.11   &0.0000  &29.29  &56.58 \\ \hline
    
    \multirow{4}{*}{\begin{tabular}[c]{@{}c@{}}\textit{Hashmasks} 
    \end{tabular}}      
    &$Q_1$ &24,229.31   &420.00  &0.0001  &1.96    &--  \\ 
    &$Q_2$ &23,855.67   &420.00  &0.0001  &2.23    &-1.54  \\ 
    &$Q_3$ &26,318.11   &420.00  &0.0000  &2.33    &10.32  \\ 
    &$Q_4$ &26,375.24   &420.00  &0.0000  &2.46    &0.22  \\ \hline

    \multirow{4}{*}{\begin{tabular}[c]{@{}c@{}}\textit{Polychain} 
    \end{tabular}}      
    &$Q_1$ &1,206.99   &25.00   &0.0001  &0.16    &-- \\
    &$Q_2$ &1,531.16   &25.00   &0.0001  &0.21    &26.86 \\
    &$Q_3$ &1,570.90   &25.00   &0.0000  &0.24    &2.60 \\
    &$Q_4$ &1,562.35   &25.00   &0.0000  &0.24    &-0.54 \\ \hline
    
    \multirow{4}{*}{\begin{tabular}[c]{@{}c@{}}\textit{NEKO} 
    \end{tabular}}   
    &$Q_1$ &24.75       &1.60    &0.0001  &0.00    &-- \\
    &$Q_2$ &302.32      &3.20    &0.0001  &0.01    &1,121.43 \\
    &$Q_3$ &1,043.40    &4.04    &0.0000  &0.23    &245.13 \\
    &$Q_4$ &1,024.57    &4.04    &0.0000  &0.22    &-1.81 \\ \hline
    
    \multirow{4}{*}{\begin{tabular}[c]{@{}c@{}}\textit{EtherThings} 
    \end{tabular}}      
    &$Q_1$ &3.18    &0.50    &0.0001  &0.0003  &-- \\
    &$Q_2$ &3.21    &0.50    &0.0001  &0.0005  &0.94 \\
    &$Q_3$ &3.71    &0.50    &0.0000  &0.0005  &15.56 \\
    &$Q_4$ &458.93  &8.81    &0.0000  &0.0166  &12,280.41 \\ \hline
    \end{tabular}
}
\label{tab:quintilestats}
\end{table}


As demonstrated in the observed series of training collections (Table~\ref{tab:quintilestats}), we present the key market statistics. The \textit{market capitalization} (mkt cap) reflects the total value of the NFT collection's outstanding tokens owned by NFT holders at a certain point in time. 
A large market capitalization indicates the popularity and recognition of a collection. It can be categorized into three tiers: Tier 1 ($mkt \ cap > 15K$), Tier 2 ($2K < mkt \ cap < 15K$), and Tier 3 ($mkt \ cap < 2K$).
\textit{High}, \textit{low}, and \textit{mean} are the transactions of tokens, where the highest, lowest values, and mean transaction values are shown. It shows the range of transactions across tokens in the same collection. As for \textit{Change (\%)}, the change of total market value from the previous quarter is taken, showing its growth or decline in percentage terms.

\vspace{1mm} 
\noindent\textbf{Evaluation Collections.}
\noindent We first evaluate the contextual generative approach on four NFT collections, \textit{Chubbies}, \textit{CryptoTrunks}, \textit{VeeFriends}, and \textit{MAYC}. These collections have the following market capitalization:
\vspace{1mm}
\begin{itemize}
\addtolength{\itemindent}{0.9cm}
\setlength{\itemsep}{3pt}%
    \item[($X_6$)] \emph{Chubbies} --- ETH 3K 
    \item[($X_7$)] \emph{CryptoTrunks} --- ETH 620 
    \item[($X_8$)] \emph{VeeFriends} --- ETH 60K 
    \item[($X_{9}$)] \emph{Mutant Ape Yacht Club (MAYC)} --- ETH 484K 
\end{itemize}
\vspace{1mm}
As these test sets are only observed in the early stage $Q_1$, we only use the $Q_1$ series for learning their contextual information, leaving the series in the growth stage for evaluation. The market statistics (see Table~\ref{tab:teststats}) of the four test collections show their transaction characteristics in $Q_1$.
    \begin{table}[!htbp]
\centering
\caption{Market statistics of the four test collections in the early stage $Q_1$.}
\resizebox{0.99\linewidth}{!}{%
    \begin{tabular}{|c|c|cccc|}
    \hline
    \textbf{Collection} & \textbf{Quarter} &\textbf{Market Cap} &\textbf{High} & \textbf{Low} & \textbf{Mean}  \\ \hline 
    
    \textit{Chubbies}       &$Q_1$ &3,123.25    &20.00   &0.0001  &0.50     \\ 
    \textit{CrytoTrunks}    &$Q_1$ &1,251.49    &10.00   &0.0001  &0.24     \\ 
    \textit{VeeFriends}     &$Q_1$ &22,311.78   &100.00  &0.0001  &1.82     \\ 
    \textit{MAYC}           &$Q_1$ &41,034.89   &410.01  &0.0001  &1.38     \\ \hline
    
    \end{tabular}
}
\label{tab:teststats}
\end{table}

\begin{figure}[!htbp]
\centering
    \begin{minipage}[c]{0.95\linewidth}
        \centering
        \includegraphics[width=\linewidth]{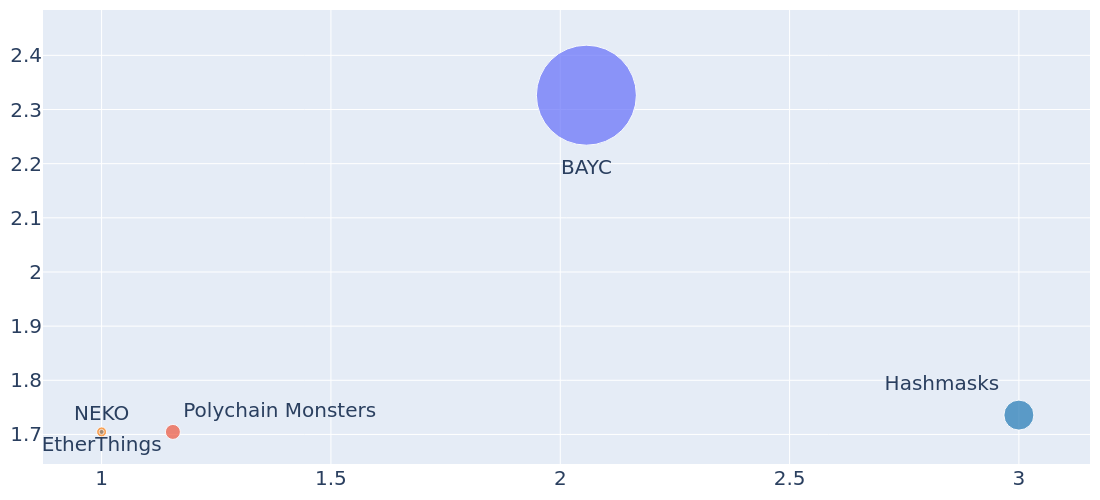}
        \caption*{(a) Training embeddings ($C_1$--$C_5$)}
    \end{minipage} 
    
    \begin{minipage}[c]{0.95\linewidth}
        \centering
        \includegraphics[width=\linewidth]{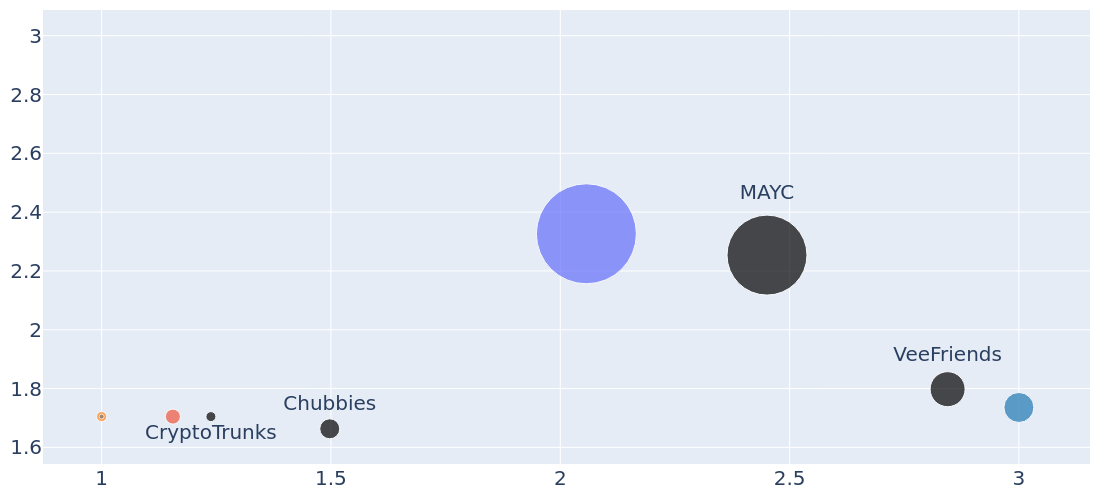}
        \caption*{(b) Test embeddings ($C_6$--$C_9$) in black}
    \end{minipage}
\caption{Training set embeddings (top) are plotted according to the market capitalization of the NFT collections. For example, BAYC in purple has the biggest point as it has the largest market value. Test embeddings (bottom) in black are overlayed with the training embeddings, showing the closeness of each test set to the respective training sets. 
}
\label{fig:embedding}
\end{figure}

\vspace{1mm} 
\noindent\textbf{Contextual Information.}
The test collection market statistics (Table~\ref{tab:teststats}) show that each test collection is more similar to a particular training collection. We can see that the statistics of CryptoTrunks are closest to NEKO, Chubbies to both Polychain and NEKO. While VeeFriends is most similar to Hashmasks, MAYC is closest to BAYC. 
Since collections are more similar to some than others, we want to make use of this information in our contextual generative approach. Hence, we take the six-dimensional PCA context vectors of each collection series, normalized between 1--3, and plot the first two dimensions (see Figure~\ref{fig:embedding}). 

In Figure~\ref{fig:embedding}a, the first two dimensions of the training collections PCA contexts are plotted according to their market capitalization sizes. As shown, the collections with similar market capitalizations and transaction characteristics are closer, while the collections with larger values are further in the 2D space.
Next, we overlay the four test collection context vectors on the training contexts (Figure~\ref{fig:embedding}b). 
As expected, we can see that VeeFriends, having a similar market size as Hashmasks, are closer in the embedding space. MAYC is closest to BAYC. The other two test collections with smaller market values, Chubbies and CryptoTrunks, are closer to the training collections around the same market size. 
Below are the entire 6D context vectors for each collection:
\begin{enumerate}
\centering
\setlength{\itemsep}{2pt}%
    \item[$C_1$:] (2.06, 2.33, 1.72, 1.81, 1.92, 1.90)
    \item[$C_2$:] (3.00, 1.74, 1.93, 1.89, 1.85, 1.82)
    \item[$C_3$:] (1.16, 1.70, 1.90, 1.88, 1.84, 1.88)
    \item[$C_4$:] (1.00, 1.70, 1.90, 1.88, 1.84, 1.88)
    \item[$C_5$:] (1.00, 1.70, 1.90, 1.88, 1.84, 1.88)
    \item[$C_6$:] (1.50, 1.66, 1.88, 1.89, 1.85, 1.85)
    \item[$C_7$:] (1.24, 1.70, 1.89, 1.87, 1.86, 1.88)
    \item[$C_8$:] (2.84, 1.80, 1.88, 1.84, 1.89, 1.82)
    \item[$C_9$:] (2.45, 2.25, 1.78, 1.79, 1.93, 1.85)
\end{enumerate}
Using these vectors as the respective contexts for each collection, we are able to provide more information to our model. These contexts allow us to distinguish transaction series among the various collections better, improving model generation.

\begin{table*}[!htbp]
\centering
\caption{Performance measures (regression statistics and market metrics) of the transaction series generated by our contextual generative models, ContextPred and NFTContextPred, and the baseline single collection models. Values that reflect the best performance are in bold. Our contextual approach mostly achieves the highest regression statistics and closely matches the actual market metrics compared to the baseline models.}
\resizebox{0.75\linewidth}{!}{%
\begin{tabular}{c|c|cccccccc|c}
    \multirow{2}{*}{\begin{tabular}[c]{@{}c@{}}\textbf{NFT Collection} \end{tabular}}
    & \multirow{2}{*}{\begin{tabular}[c]{@{}c@{}}\textbf{Model} \end{tabular}}
    & \multicolumn{4}{c}{\textbf{Regression Stats}} & \multicolumn{4}{c}{\textbf{Market Cap}} & \textbf{Tier} \\
    & &\textbf{MAE} & \textbf{MSE} & \textbf{RMSE} & $\mathbf{R^2}$ & $\mathbf{Q_1}$ & $\mathbf{Q_2}$ & $\mathbf{Q_3}$ & $\mathbf{Q_4}$ & \\ \hline 

    \multirow{8}{*}{\begin{tabular}[c]{@{}c@{}}\textbf{\textit{Chubbies}} \end{tabular}} 
        \rule{0pt}{3ex}   &Actual  &--&--&--&-- &3,132.28 &3,049.97 &2,967.42 &2,917.50 &2   \\ \cline{2-11}
        \rule{0pt}{3ex}   &$\mathcal{M}_{1}$   &46.80 &7,105.05 &84.29 &-5,132.98 &0.00 &22.96 &119.81 &579.10 &1   \\
        \rule{0pt}{3ex}   &$\mathcal{M}_{2}$   &0.41   &0.80 &0.89 &0.36 &0.00 &0.26  &0.29   &0.31 &\underline{2}   \\
        \rule{0pt}{3ex}   &$\mathcal{M}_{3}$   &0.26 &0.62  &0.79 &0.51 &0.00 &0.18  &0.27   &0.33 &3   \\
        \rule{0pt}{3ex}   &$\mathcal{M}_{4}$   &0.33 &1.07  &1.03 &0.14 &0.00 &0.87  &0.86   &0.86 &3   \\
        \rule{0pt}{3ex}   &$\mathcal{M}_{5}$   &0.35 &1.09  &1.05 &0.11 &0.00 &0.95  &0.95   &0.95 &3   \\
        \rule{0pt}{3ex}   &$\mathcal{M}_{\bf{X}}$  &0.35 &0.57  &0.76 &0.55 &0.00 &0.99  &0.99   &0.99 &3   \\
        \rule{0pt}{3ex}   &\emph{ContextPred} &0.32 &0.69  &0.83 &0.45 &0.00 &0.27  &0.25   &0.23 &\underline{2}   \\ 
        \rule{0pt}{3ex}   &\emph{NFT ContextPred}  &\bf{0.18} &\bf{0.23}  &\bf{0.48} &\bf{0.83} &0.00 &\bf{0.07}  &\bf{0.07}   &\bf{0.05} &\underline{2}   \\ \hline 
        
    \multirow{8}{*}{\begin{tabular}[c]{@{}c@{}}\textbf{\textit{CryptoTrunks}}  \end{tabular}} 
        \rule{0pt}{3ex}   &Actual  &--&--&--&-- &1,250.64 &1,381.67 &1,383.83 &1,381.40 &3   \\ \cline{2-11}
        \rule{0pt}{3ex}   &$\mathcal{M}_{1}$   &46.29   &6,914.36 &83.15 &-11,045.39 &0.00 &30.18 &152.28 &726.26 &1   \\
        \rule{0pt}{3ex}   &$\mathcal{M}_{2}$   &0.38   &0.45 &0.67 &0.28 &0.00 &0.65  &0.64   &0.63 &2   \\
        \rule{0pt}{3ex}   &$\mathcal{M}_{3}$   &0.27 &\bf{0.34}  &\bf{0.58} &\bf{0.44} &0.00 &0.13  &0.17   &0.20 &\underline{3}   \\
        \rule{0pt}{3ex}   &$\mathcal{M}_{4}$   &0.26 &0.53  &0.73 &0.13 &0.00 &0.82  &0.82   &0.82 &\underline{3}   \\
        \rule{0pt}{3ex}   &$\mathcal{M}_{5}$   &0.27 &0.55  &0.74 &0.10 &0.00 &0.94  &0.94   &0.94 &\underline{3}   \\
        \rule{0pt}{3ex}   &$\mathcal{M}_{\bf{X}}$  &0.32 &0.52  &0.72 &0.16 &0.00 &1.18  &1.18   &1.17 &\underline{3}   \\
        \rule{0pt}{3ex}   &\emph{ContextPred}  &0.31 &0.43  &0.65 &0.31 &0.00 &\bf{0.04}  &\bf{0.04}   &\bf{0.03} &\underline{3}   \\ 
        \rule{0pt}{3ex}   &\emph{NFT CondPred}  &\bf{0.22} &0.35  &0.59 &\bf{0.44} &0.00 &0.13  &0.13   &0.14 &\underline{3}   \\ \hline
        
    \multirow{8}{*}{\begin{tabular}[c]{@{}c@{}}\textbf{\textit{VeeFriends}}  \end{tabular}} 
        \rule{0pt}{3ex}   &Actual  &--&--&--&-- &22,287.76 &32,894.39 &37,948.39 &43,902.90 &1   \\ \cline{2-11}
        \rule{0pt}{3ex}   &$\mathcal{M}_{1}$   &53.78   &9,944.38   &99.72   &-190.46 &0.00 &3.31  &16.12  &61.27 &\underline{1}   \\
        \rule{0pt}{3ex}   &$\mathcal{M}_{2}$   &2.37   &49.04 &7.00 &-0.21 &0.00 &0.78  &0.84   &0.86 &2   \\
        \rule{0pt}{3ex}   &$\mathcal{M}_{3}$   &2.28 &34.26 &5.85 &0.12 &0.00 &0.72  &0.83   &0.91 &2   \\
        \rule{0pt}{3ex}   &$\mathcal{M}_{4}$   &2.88 &39.74 &6.30 &-0.05 &0.00 &0.98  &0.98   &0.99 &3   \\
        \rule{0pt}{3ex}   &$\mathcal{M}_{5}$   &2.91 &39.99 &6.32 &-0.06 &0.00 &0.99  &0.99   &1.00 &3   \\
        \rule{0pt}{3ex}   &$\mathcal{M}_{\bf{X}}$  &2.62 &33.72 &5.81 &0.16 &0.00 &0.92  &0.93   &0.94 &2   \\
        \rule{0pt}{3ex}   &\emph{ContextPred}  &2.24 &30.33 &5.51 &0.25 &0.00 &0.74  &0.77   &0.80 &\underline{1}   \\ 
        \rule{0pt}{3ex}   &\emph{NFT ContextPred}  &\bf{1.56} &\bf{24.45} &\bf{4.94} &\bf{0.40} &0.00 &\bf{0.49}  &\bf{0.57}   &\bf{0.62} &\underline{1}   \\ \hline         
        

    \multirow{8}{*}{\begin{tabular}[c]{@{}c@{}}\textbf{\textit{MAYC}}  \end{tabular}} 
        \rule{0pt}{3ex}   &Actual  &--&--&--&-- &41,034.89 &125,819.99 &208,483.79 &214,059.32 &\underline{1}   \\ \cline{2-11}
        \rule{0pt}{3ex}   &$\mathcal{M}_{1}$   &52.64   &10,310.33  &101.54  &-23.07 &0.00  &1.03  &4.51  &19.63  &1    \\
        \rule{0pt}{3ex}   &$\mathcal{M}_{2}$   &8.00    &394.76     &19.87   &-0.08  &0.00  &0.86  &0.92  &0.92   &2    \\
        \rule{0pt}{3ex}   &$\mathcal{M}_{3}$   &7.78    &338.24     &18.39   &0.03   &0.00  &0.89  &0.95  &0.97   &2    \\
        \rule{0pt}{3ex}   &$\mathcal{M}_{4}$   &8.27    &351.10     &18.74   &-0.03  &0.00  &0.99  &1.00  &1.00   &3    \\
        \rule{0pt}{3ex}   &$\mathcal{M}_{5}$   &8.28    &351.81     &18.76   &-0.03  &0.00  &1.00  &1.00  &1.00   &3    \\
        \rule{0pt}{3ex}   &$\mathcal{M}_{\bf{X}}$  &7.49    &428.72     &20.71   &-0.14  &0.00  &0.83  &0.89  &0.90   &\underline{1}   \\
        \rule{0pt}{3ex}   &\emph{ContextPred}  &7.71    &311.78     &17.66   &0.16   &0.00  &0.84  &0.90  &0.91   &\underline{1}   \\ 
        \rule{0pt}{3ex}   &\emph{NFT ContextPred}  &\bf{6.71}    &\bf{284.03}     &\bf{16.85}   &\bf{0.24}   &0.00  &\bf{0.72}  &\bf{0.83}  &\bf{0.84}   &\underline{1}   \\ \hline         
        
\end{tabular}
}
\label{tab:predictperf}
\end{table*} 
    
\subsection{Performance}
Next, we evaluate our approach to generating NFT transactions (see Table~\ref{tab:predictperf}). 
%
%
For the evaluation baselines, models $\mathcal{M}_{1}$, $\mathcal{M}_{2}$, $\mathcal{M}_{3}$, $\mathcal{M}_{4}$, and $\mathcal{M}_{5}$ are trained only on collections $X_1, \dots, X_5$, respectively. Next, we train another model $\mathcal{M}_{\bf{X}}$ on all five collections $X_1$--$X_5$ for comparison. Finally, we compare our proposed methods, \emph{ContextPred} (contextual-aware) and \emph{NFT ContextPred} (contextual-aware $+$ step-transform), with the baseline models. 

The proposed approach consistently generates transactions with values closer to the regression statistics~\footnote{Statistics used to measure the regression analysis include MAE (mean absolute error), MSE (mean squared error), RMSE (root mean square error), and $R^2$ (R-Squared, the coefficient of determination).} and actual market metrics than other models, projecting the future market value of NFTs. We take the absolute difference percentage as our market value generation performance metrics. It is computed by: 
\begin{equation*}
\frac{\mid y-\hat{y} \mid}{y}
\end{equation*}
where $y$ is the actual market value in any particular quarter and $\hat{y}$ the model generated value.

Given that only the $Q_1$ transactions of the test NFTs are observed, our approach first computes the context vectors of the test collections. Using these contexts as additional information, the approach then generates the series of quarters $Q_2$, $Q_3$, and $Q_4$. We present the evaluation metrics and results.
A higher similarity (smaller differences) indicates a greater resemblance to the real future series, achieving better generation performance that effectively captures the transaction network characteristics of the future series and projects the potential market value of NFTs.

As demonstrated in the results, the baseline model $\mathcal{M}_{2}$ trained on Hashmasks ($X_2$) has the generated transactions closest to VeeFriends ($X_8$) as they are closest in market values, as shown in their contexts (Figure~\ref{fig:embedding}). Similarly, baseline $\mathcal{M}_{3}$ performs best in projecting $X_6$ and $X_7$ due to their similarities, and $\mathcal{M}_{1}$ generates transactions most similar to $X_9$. However, these baselines perform poorly on test sets that are different. The baseline model $\mathcal{M}_{\bf{X}}$ trained on all $X_1$--$X_5$ performs poorly for all as it is unable to differentiate between series from various collections.

Whereas, we can see that \emph{ContextPred} performs better than most of the individual baselines on all test sets. It shows that the context vectors provide the aggregate model with essential identifying information while leveraging a large amount of available data from various collections. 
On top of that, \emph{NFT ContextPred} achieves even better performance than our contextual model, demonstrating the effectiveness of our step-transform function in characterizing NFT transactions.

\subsection{Analysis}
As the characteristics of the NFT token transaction series are piecewise constant, the aim of our approach, in addition to generating transactions of market value, is to model these step functions faithfully.
    \begin{figure}[H] 
    \centering
    \includegraphics[width=0.8\linewidth]{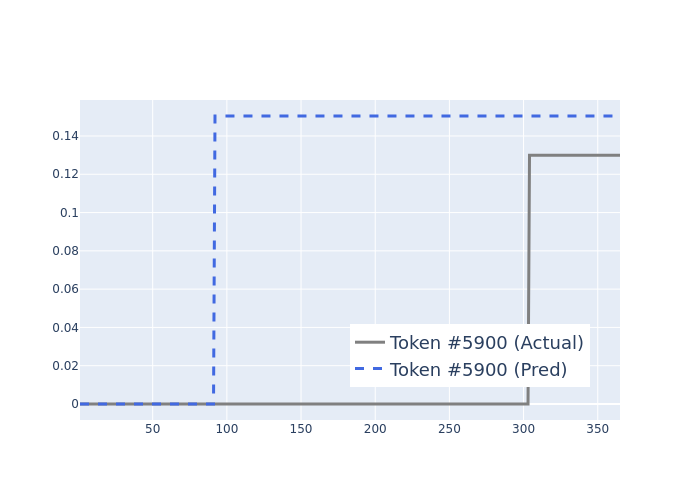}
    \caption{Sample token \#5900 from test collection Chubbies, showing the actual and generated series. 
    }
    \label{fig:stepx6}
\end{figure}    

We sample a token generation contrasted with the actual series from the VeeFriends collection (see Figure~\ref{fig:stepx6}), showing that our approach generates transactions following the characteristics of actual NFT transaction series (see Appendix for samples from other test collections).

\begin{figure}[!htbp]
\centering
    
    \begin{minipage}[c]{0.95\linewidth}
        \centering
        \includegraphics[width=\linewidth]{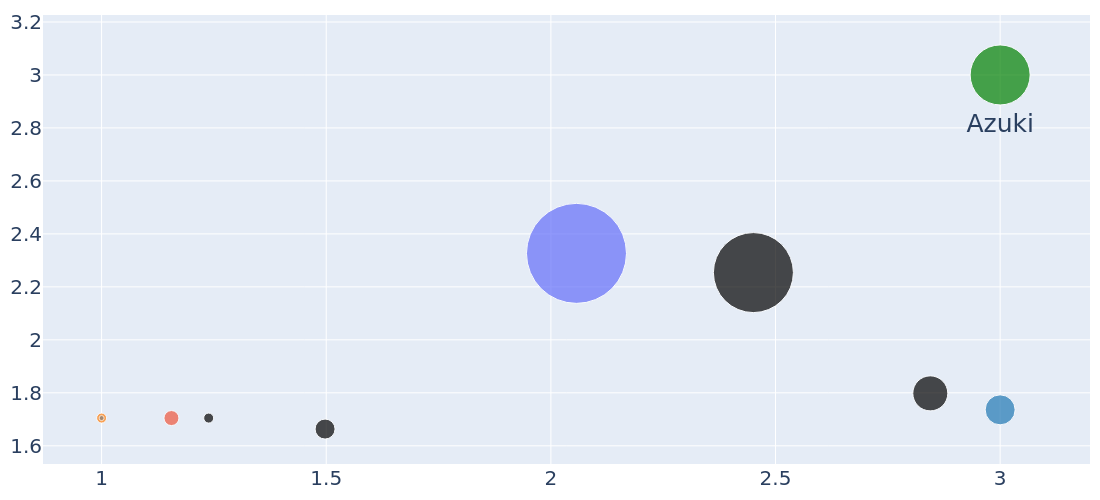}
    \end{minipage}
\caption{The contextual embedding of the Azuki collection in green among all embeddings, showing its distance from the other collections.
}
\label{fig:emb_azuki}
\end{figure}

\vspace{1mm} 
\noindent\textbf{Distant Contexts.}
In addition, we evaluate another collection \textit{Azuki}. It is one of the latest top NFT collections containing 10,000 tokens that was minted on Jan 12, 2022, at the price of 1 ETH. 
Azuki has had a meteoric rise in popularity, reaching a total transaction volume of \textit{ETH 274K} in a year. Due to the unique transactions of this collection, its contextual vector is far in the embedding space from the training embedding vectors that were available to our contextual generative methods (see Figure~\ref{fig:emb_azuki}). 

\noindent As such, using the Azuki collection context vector: 
\begin{equation*}
    (3.00, 3.00, 2.36, 1.36, 2.21, 1.66)    
\end{equation*}
in our contextual generative approach, caused the performance to drop below most other uncontextualized methods (see Table~\ref{tab:azuki_analysis}). 
\begin{table}[!htbp]
\centering
\caption{Generative performance on the \textit{Azuki} collection.}
\resizebox{1.0\linewidth}{!}{%
\begin{tabular}{c|cccccccc}
    \multirow{2}{*}{\begin{tabular}[c]{@{}c@{}}\textbf{Model} \end{tabular}}
    & \multicolumn{4}{c}{\textbf{Regression Stats}} & \multicolumn{4}{c}{\textbf{Market Cap}} \\ 
    &\textbf{MAE} & \textbf{MSE} & \textbf{RMSE} & $\mathbf{R^2}$ & $\mathbf{Q_1}$ & $\mathbf{Q_2}$ & $\mathbf{Q_3}$ & $\mathbf{Q_4}$ \\ \hline 

        \rule{0pt}{3ex}   Actual  &--&--&--&-- &111,291.17  &130,685.38  &130,602.78  &138,343.15    \\ \cline{1-9}
        \rule{0pt}{3ex}   $\mathcal{M}_{1}$   &103.26 &34,072.93 &184.59  &-92.61 &0.00 &2.76 &11.95 &28.57    \\
        \rule{0pt}{3ex}   $\mathcal{M}_{2}$   &10.44  &1,425.86  &37.76   &-3.03  &0.00 &0.42 &1.12  &1.11    \\
        \rule{0pt}{3ex}   $\mathcal{M}_{3}$   &9.24   &373.01    &19.31   &-0.07  &0.00 &0.83 &1.50  &1.50    \\
        \rule{0pt}{3ex}   $\mathcal{M}_{4}$   &10.33  &404.79    &20.12   &-0.16  &0.00 &0.96 &1.59  &1.54    \\
        \rule{0pt}{3ex}   $\mathcal{M}_{5}$   &10.36  &405.68    &20.14   &-0.16  &0.00 &0.96 &1.59  &1.55    \\
        \rule{0pt}{3ex}   $\mathcal{M}_{\bf{X}}$  &7.46   &595.69    &24.41   &-0.67  &0.00 &0.62 &1.23  &1.18     \\
        \rule{0pt}{3ex}   \emph{ContextPred}  &58.00  &4,885.68  &69.90   &-12.87 &0.00 &6.15 &5.55  &5.20     \\ 
        \rule{0pt}{3ex}   \emph{NFT ContextPred}  &52.17  &4,206.39  &64.86   &-10.93 &0.00 &5.84 &5.21  &4.88     \\ \hline

\end{tabular}
}
\label{tab:azuki_analysis}
\end{table} 
    
\noindent These contexts play a significant role in the contextual generative approach. Hence, it is essential to (1) obtain relevant contextual information characteristic of the collections and (2) train the model with varying contexts, ensuring that it is extensive enough to include diverse characteristics.

\vspace{1mm}
\section{Conclusion}
\label{sec:conclude}
NFT collections are difficult to project. In this paper, we approached the challenge by studying the characteristics of NFT transactions in the various stages of their development. We proposed a two-step contextual generative approach that first extracts the contextual information of NFT collections from their early transactions using unsupervised learning. Next, the approach takes in these contexts and early transactions, progressively generating future token transactions of a newly minted collection. 
Our evaluation results demonstrate the two-step approach's generative power on the underlying transactions of NFT collections. It opens up an exciting field of exploration that potentially increases NFT market projection capabilities. Our current contextual generative modeling approach is conditioned on linear transaction value correlation information. As future work, changes through the development stages could be considered to generate adaptive generations. Our work acts as a stepping stone for future contextual approaches to NFT transaction generation and other applications.

\newpage
\bibliographystyle{ACM-Reference-Format}
\bibliography{reference}

\appendix
\section*{Appendix} 

\subsection*{A. Transaction Analysis}
We perform additional analysis on the transaction, where the transaction counts of each token in the first year are aggregated and plotted in histograms. The reported numbers show the typical number of transactions in a year of each collection.

\begin{figure}[H] 
    \centering
    \includegraphics[width=0.66\linewidth]{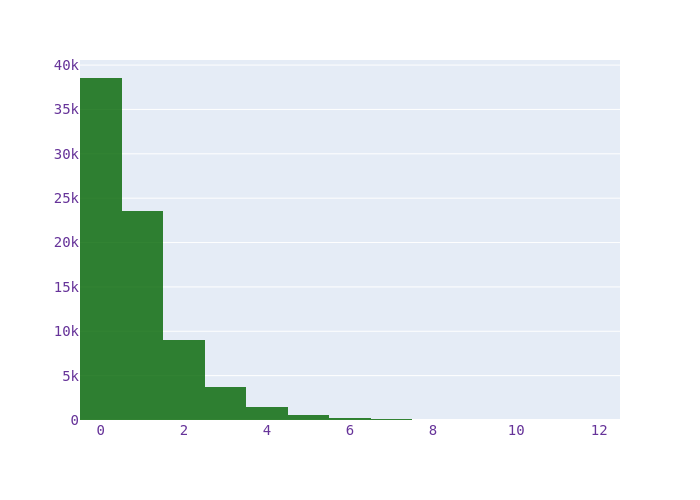}
    \caption{Most tokens from the five training NFT collections 
    are not traded in the first year, while the majority of traded tokens are transacted 2 or 3 times. 
    }
    \label{fig:translen_train}
\end{figure}

\vspace{3mm} 
\noindent\textbf{All training collections.}
In the following analysis, the first histogram consists of the number of transactions (x-axis) and the token counts (y-axis) from all five training collections: BAYC, Hashmasks, Polychain Monsters, NEKO, and EtherThings. Most tokens are not transacted in the first year (zero transactions), and tokens that are traded have only 2 or 3 transactions (see Figure~\ref{fig:translen_train}).

\noindent\textbf{Top two training collections.}
However, the transactions of the top two collections in the training set are very different from the aggregate transactions (see Figure~\ref{fig:translen_traintop2}). Most of the tokens in these two collections are at least traded once, and most of them have 2--4 transactions, as shown below:
\begin{figure}[H] 
\centering
    \begin{minipage}[c]{0.55\linewidth}
        \centering
        \includegraphics[width=\linewidth]{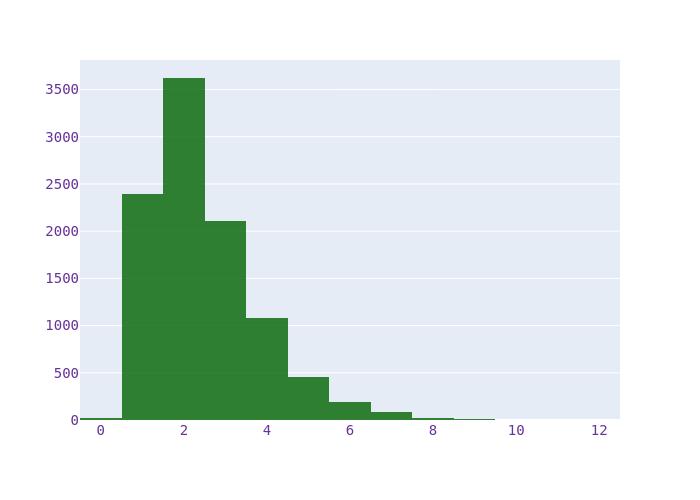}
    \end{minipage} 
    
    \begin{minipage}[c]{0.55\linewidth}
        \centering
        \includegraphics[width=\linewidth]{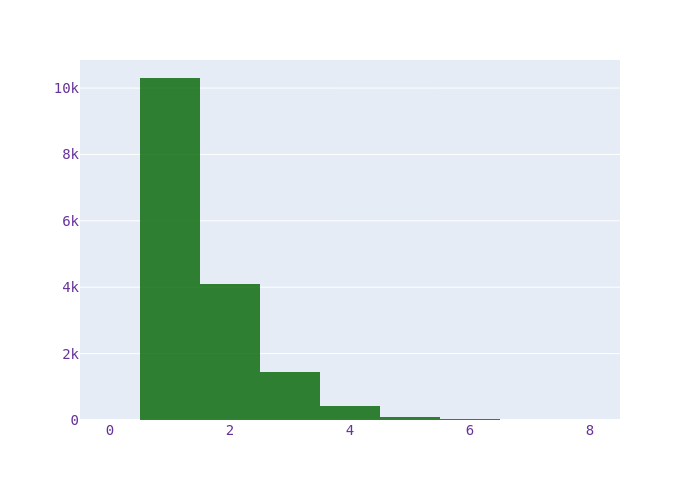}
    \end{minipage}
\caption{Transactions of the top two collections, BAYC (top) and Hashmasks (bottom), in the training set. 
Most tokens are traded 2--4 times, in the first year.
}
\label{fig:translen_traintop2}
\end{figure}

\begin{figure}[H] 
\centering
    \begin{minipage}[c]{0.55\linewidth}
        \centering
        \includegraphics[width=\linewidth]{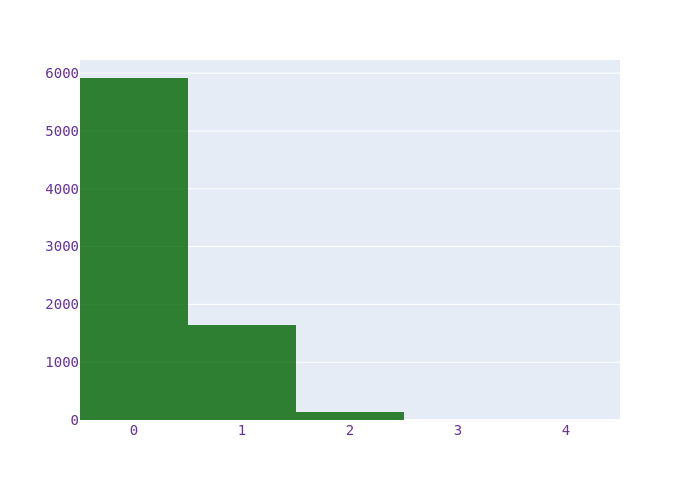}
    \end{minipage} 
    
    \begin{minipage}[c]{0.55\linewidth}
        \centering
        \includegraphics[width=\linewidth]{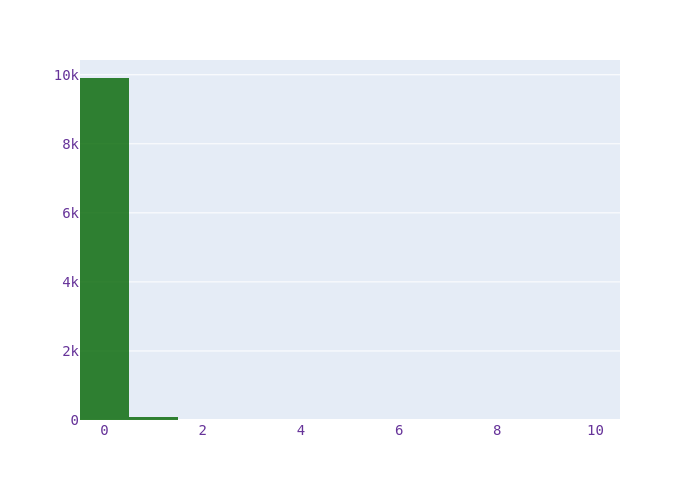}
    \end{minipage}
\caption{Transactions of the bottom two collections, NEKO (top) and EtherThings (bottom), in the training set. Most of the tokens are not even traded in the first year, with tokens that are only traded once, if at all. 
}
\label{fig:translen_trainbottom2}
\end{figure}

\noindent\textbf{Bottom two training collections.}
On the contrary, the bottom two collections in the same training set have markedly different token transactions. As demonstrated in the histograms below, most of the tokens were never traded in the first year, resulting in zero transactions (see Figure~\ref{fig:translen_trainbottom2}).
Hence, the transactions vary widely across various collections in the training set.

\subsection*{B. Implementation Details}
The basesline models are based on a stacked 2-layer LSTM architecture. The input to the models is a batch of transaction series length 20 and dimension 2, where the batch size is 1024. We select the series length as 20 because it captures the structures of most transactions of different collections. The LSTM hidden layers are designed with 300 memory units, which is more than sufficient to learn the transaction series of NFT collections.
The model architecture is listed in Table~\ref{tab:hypers}.
    \begin{table}[H] 
\centering
\caption{Model architecture configuration. Hyperparameters of the baseline models (unconditional) and the contextual LSTM generative model (conditional) for NFT projection.}
\resizebox{0.99\linewidth}{!}{%
\begin{tabular}{|c|c|c|}
    \hline
    & Unconditional LSTM & Conditional LSTM \\ \hline
    Input           &(20, 300, 1) & (26, 300, 1)   \\ \hline
    \nth{1} layer   &300, LSTM, ReLU &300, LSTM, Sigmoid     \\ \hline
    \nth{2} layer   &300, LSTM, ReLU &300, LSTM, Sigmoid     \\ \hline
    \nth{3} layer   &(1, 2), fully-connected, linear &(1, 2), fully-connected, linear     \\ \hline
    Dropout rate    &0.2 &0.2     \\ \hline 
    Epoch           &50 &50       \\ \hline 
    Batch size      &1024 &1024   \\ \hline 
    Optimizer       &Adam &Adam   \\ \hline 
    Loss            &Mean Squared Error (MSE) &Mean Squared Error (MSE)   \\ \hline 
\end{tabular}
}
\label{tab:hypers}
\end{table}

As for the contexts, we use a 6D condition vector for both the \emph{ContextPred} and \emph{NFT ContextPred} models, resulting in an input length of 26. However, the rest of the architecture remains the same, such that it is a fair comparison.

\subsection*{C. Token Samples}
\begin{figure}[H] 
    \centering
    \includegraphics[width=0.8\linewidth]{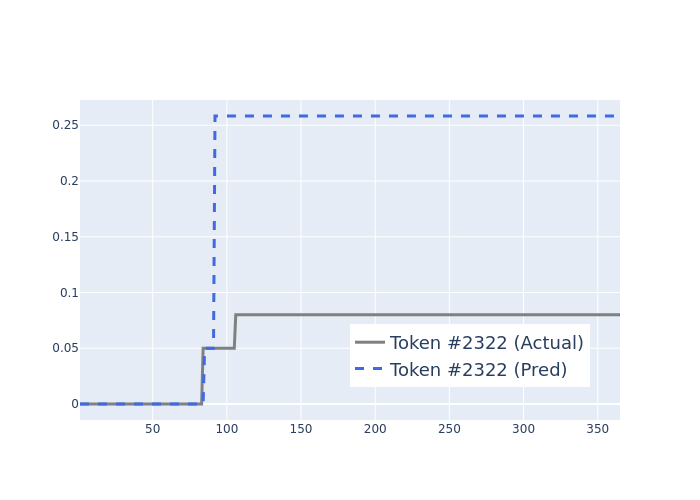}
    \caption{Sample token \#2322 from test collection CryptoTrunks, showing the actual and generated series. 
    }
    \label{fig:stepx7}
\end{figure}

\begin{figure}[H] 
    \centering
    \includegraphics[width=0.8\linewidth]{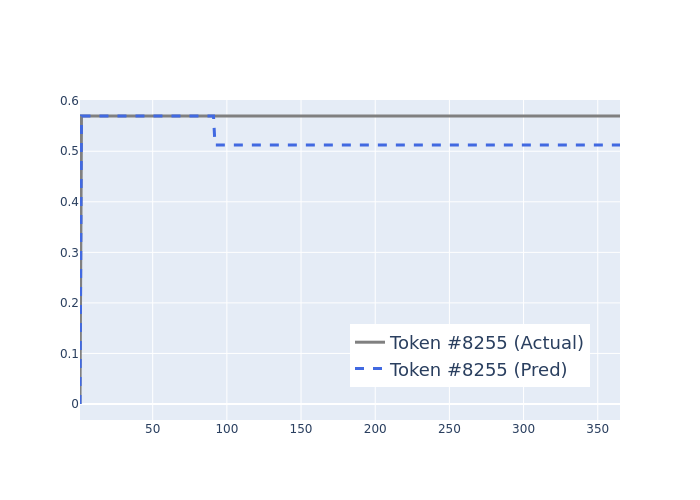}
    \caption{Sample token \#8255 from test collection VeeFriends, showing the actual and generated transaction series. 
    }
    \label{fig:stepx8}
\end{figure}

From the model \emph{NFT ContextPred} generation of test collections CryptoTrunks and VeeFriends, we sample a token from each collection, respectively. These samples (see Figures \ref{fig:stepx7} and \ref{fig:stepx8}) demonstrate the piecewise constant functions following the NFT transaction characteristics.

\subsection*{D. Description of NFT Collections}

\noindent\textbf{Training sets.}
\noindent There are five NFT collections used for training. Description below:
\begin{itemize}
    \item \emph{Bored Ape Yacht Club (BAYC).} 
    Minted on Apr 23, 2021.
    The Bored Ape Yacht Club NFT collection consists of 10,000 unique Bored Ape NFT tokens. These tokens are unique digital collectibles on the Ethereum blockchain.
    Moreover, the tokens also double as membership to the Yacht Club (\url{www.BoredApeYachtClub.com}), granting access to benefits such as a collaborative graffiti board where only members are allowed. 

    \item \emph{Hashmasks.} Minted on Jan 28, 2021. This NFT collection consists of 16,384 unique NFTs created by 70 artists globally. It is a digital art collection that uniquely allows the buyer of a token to name it, gaining its rarity by the name that the consumer instead of the artist gives. Each token presents a human-like figure. They are designed to be bizarre,  eccentric, and ultimately intriguing. While the tokens resemble humans, some are extraterrestrial looking and carry an odd combination of traits such as hairstyle, eye color, and clothing.

    \item \emph{Polychain Monsters.} Minted on Mar 31, 2021. At its core, Polychain monsters are beautifully animated NFTs, digital collections with varying degrees of scarcity.
    
    \item \emph{NEKO Official.} Minted on Apr 18, 2021. NEKOs are 10,000 uniquely generated animated cat NFTs on the Ethereum blockchain, where no two NFTs are identical.

    \item \emph{EtherThings.} Minted on Apr 14, 2021. EtherThings is an NFT Collection on the Ethereum blockchain with a limited supply of 10,000 tokens. EtherThings is inspired by human faces. The tokens are designed to express the diversity of human characteristics, representing the pursuit of self-discovery on the blockchain. It is a community celebrating representation, inclusivity, and diversity in web3.
\end{itemize}

\vspace{1mm} 
\noindent\textbf{Test sets.}
\noindent There are five NFT collections used for evaluation. Description below:
\begin{itemize}
    \item \emph{Chubbies.} Chubbies are NFTs on the Ethereum blockchain that are generated to be cute and chubby. Each token is a unique and programmatically generated 32x32 GIF enlarged to 320x320 pixels. They are animated with five different frames at 150ms per frame with more than six different properties. There are Zombies, Apes, Aliens, Robots, Cats, and Monkeys. 
    The collection has a limited supply of 10,000 NFT tokens. 
    
    \item \emph{CryptoTrunks.} CryptoTrunks are generated NFT trees that use oracles to create a specialized NFT. It lets users mint them to display their guilty of harming the environment through the crypto wallet's environmental impact. There is no limit on the number of minted trees.
    
    \item \emph{VeeFriends.} It is an NFT project started by an internet personality to build a community around his creative and business passions. Each NFT will have different levels of access to his content and activities. 
    
    \item \emph{Mutant Ape Yacht Club (MAYC).} Minted on Aug 28, 2021. This is a collection of mutant apes containing 20,000 tokens that can only be created by exposing an existing Bored Ape to a vial of ``MUTANT SERUM'' or by minting a Mutant Ape in the public sale.

    \item \emph{Azuki.} Minted on Jan 12, 2022. A collection of 10,000 avatars that give members access to a community of artists, builders, and web3 enthusiasts. It is a place where they meet to create a decentralized future.
\end{itemize}

\end{document}